\documentclass[%
 reprint,
 amsmath,amssymb,
 aps,
]{revtex4-2}

\usepackage{graphicx}
\usepackage{dcolumn}
\usepackage{bm}
\usepackage{hyperref}

\usepackage[dvipsnames]{xcolor}

\begin{document}

\preprint{APS/123-QED}

\title{Benchmarking of Different Optimizers in the Variational Quantum Algorithms for Applications in Quantum Chemistry}

\author{Harshdeep Singh}
\email{harshdeeps@kgpian.iitkgp.ac.in}
\affiliation{%
 Center of Computational and Data Sciences, Indian Institute of Technology, Kharagpur, India
}%
\author{Sonjoy Majumder}%
\email{sonjoym@phy.iitkgp.ac.in}
\affiliation{%
  Department of Physics, Indian Institute of Technology, Kharagpur, India}%
\author{Sabyashachi Mishra}%
\email{mishra@chem.iitkgp.ac.in}
\affiliation{%
  Department of Chemistry, Indian Institute of Technology, Kharagpur, India}%

\date{\today}

\begin{abstract}
\noindent Classical optimizers play a crucial role in determining the accuracy and convergence of variational quantum algorithms; leading algorithms use a near-term quantum computer to solve the ground-state properties of molecules, simulate dynamics of different quantum systems, and so on.  In literature, many optimizers, each having its own architecture, have been employed expediently for different applications. In this work, we consider a few popular and efficacious optimizers and assess their performance in variational quantum algorithms for applications in quantum chemistry in a realistic noisy setting.
We benchmark the optimizers with critical analysis  based on quantum simulations of simple molecules, such as Hydrogen, Lithium Hydride, Beryllium Hydride, water, and Hydrogen Fluoride. The errors in the ground-state energy, dissociation energy, and dipole moment are the parameters used as yardsticks. All the simulations were carried out with an ideal quantum circuit simulator, a noisy quantum circuit simulator, and finally, a noisy simulator with noise embedded from the IBM Cairo quantum device to understand the performance of the classical optimizers in ideal and realistic quantum environments. We used the standard unitary coupled cluster (UCC) ansatz for simulations, and the number of qubits varied from two, starting from the Hydrogen molecule to ten qubits, in Hydrogen Fluoride. Based on the performance of these optimizers in the ideal quantum circuits, the conjugate gradient (CG), limited-memory Broyden-Fletcher-Goldfarb-Shanno bound (L\_BFGS\_B), and sequential least squares programming (SLSQP) optimizers are found to be the best performing gradient-based optimizers. While constrained optimization by linear approximation (COBYLA) and POWELL perform most efficiently among the gradient-free methods. However, in noisy quantum circuit conditions, Simultaneous Perturbation Stochastic Approximation (SPSA), POWELL, and COBYLA are among the best-performing optimizers. 
\end{abstract}

\maketitle


\section{\label{sec:level1}Introduction}

\begin{figure*}
\includegraphics[width=15cm]{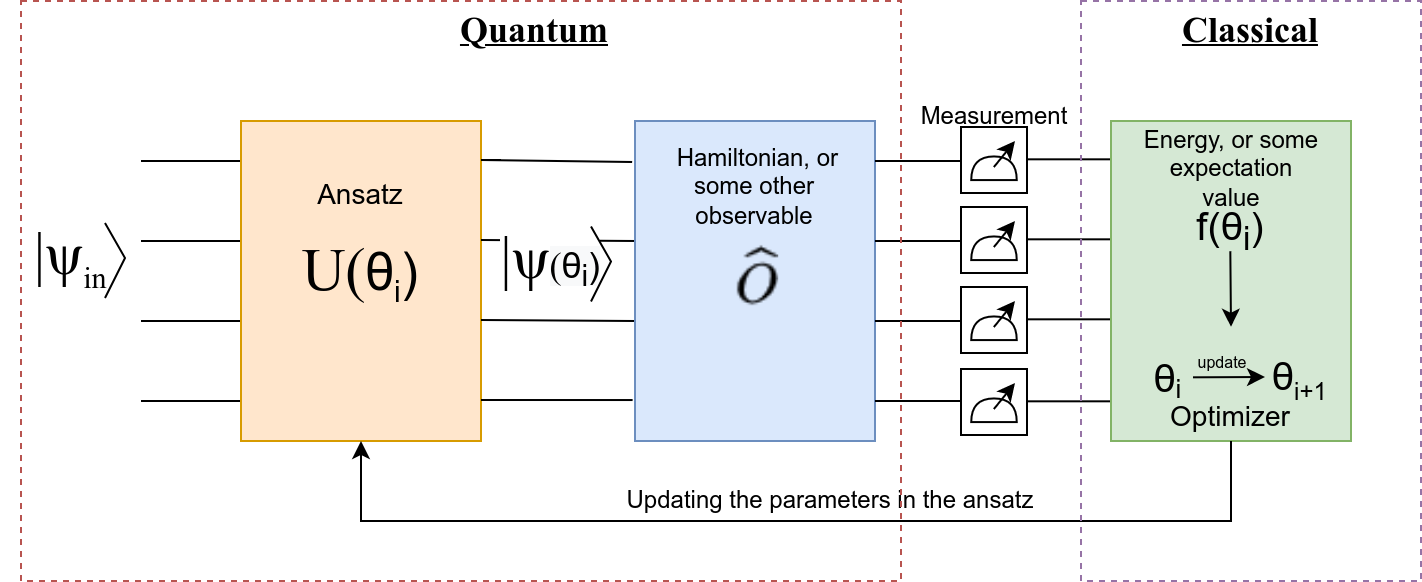}
\caption{\label{fig:vqa}Basic Structure of a Variational Quantum Algorithm: The algorithm starts with an initial state $|\Psi_{\rm in}\rangle$, which is generally obtained by setting all qubits as $|0\rangle$ state. Then, via an \textit{ansatz}, $U(\theta)$, the variational parameters are introduced into the state, and we obtain $\Psi(\theta)$. A measurement is made in this state to obtain desired observable, such as the energy of a molecular system. Everything up to this point encompasses the quantum part of the algorithm. This measurement yields the expectation value of the observable. An optimizer is used to find a better set of parameters to evaluate the optimized value of the observable in the quantum circuit. A classical algorithm is used for this part of the computation.}
\end{figure*}

\setcounter{equation}{0}
\noindent Quantum computation is defined as the design of computational methods and algorithms based on quantum mechanical principles rather than classical methods. In the last few years, 
quantum computation has emerged as one of the most popular and promising areas of research, with applications in natural sciences~\cite{first, fedorov, Protein}, machine learning~\cite{ml1,ml2}, finance~\cite{finance}, and cryptography~\cite{cryp}. Even early quantum processors like the Google Sycamore~\cite{quantumSupermacy} and some other recent experiments~\cite{super1, super2}  have shown an early glimpse of the power of quantum computation. These early promising results profoundly demonstrate the potential advantage of using quantum computation over the classical methods and the reason for the meteoric rise of interest in quantum computation.\\
\noindent The major challenge facing the emergence of `quantum supremacy' is the limited availability of quantum resources. To mitigate the shortcoming, quantum algorithms have been developed based on the available quantum resources, albeit with partial support from classical computers. These algorithms are known as variational quantum algorithms (VQA). Studies show that these algorithms can be used in a plethora of problems~\cite{ma_2020_quantum,NonLinear}, and in some cases, they are found superior to their classical counterparts~\cite{Advantage1}. VQAs take the help of a classical optimizer to train a parameterized quantum circuit~\cite{vqa}.  The basic structure of a variational quantum algorithm is depicted in FIG.~\ref{fig:vqa}.\\

\noindent There are three different components of the algorithm: the \textit{cost function}, the \textit{Ansatz}, and the \textit{optimizer}. The cost function is defined as a map from the trainable parameters $\mathbf{\theta}$ to real numbers. Common examples of the function are the error function or the energy functional of the system in the case of quantum chemistry problems. \textit{Ansatz} comprises a set of rotation gates and entanglers employed to create superposed states from the initial qubit states. Ansatz is where the variational parameters are introduced into the quantum state. Once a quantum state is created and the corresponding cost function is evaluated, we can  further improve them by training the parameters. A classical optimizer is then employed to update the parameters of the quantum circuit until a desired level of convergence is met. The optimizer makes up the \textit{classical} portion, while the rest makes up the \textit{quantum} portion of the algorithm.

\noindent In a typical VQA exercise, the first step is the preparation of the initial state, often taken as the default configuration of the qubits, i.e.,
 $$\displaystyle |\Psi_{\rm in}\rangle = |0 \cdots 0\rangle.$$ \noindent The variational parameters ($\theta_i$)
are then introduced into the circuit via the \textit{ansatz} that comprises various rotation and entanglement gates designated as $U(\theta_i)$. For a given set of variational parameters, the state of the system is given by 
$$\displaystyle |\Psi(\theta_i)\rangle = U(\theta_i) |\Psi_{\rm in}\rangle.$$ 
\noindent The energy (or any other observable) of the system in the given state is obtained by a qubit measurement of the corresponding operator ($H_{\rm qubit}$ in case of energy) and is expressed as, 
\begin{equation}
        E(\theta_i) = \langle \Psi(\theta_i)|H_{\rm qubit}|\Psi(\theta_i)\rangle.
        \label{eq:1}
\end{equation}
The variational parameters ($\theta_{i}$) are further trained by using a suitable classical optimizer to obtain an improved set of variational parameters ($\theta_{i+1}$) iteratively until convergence.\\

\noindent An optimizer plays one of the most crucial parts in any variational quantum algorithm, as it determines the overall efficiency of the employed algorithm, both in terms of accuracy and convergence speed. The performance of an optimizer depends partly on the quantum hardware (or the simulator being used) and partly on the problem at hand. Currently, a wide variety of optimizers are available for end users. They can be broadly classified into the following three categories: a) \textit{gradient-based optimizers} require evaluation of the gradient of the cost function for optimization, b) \textit{gradient-free optimizers} do not require the gradient evaluation, and c) \textit{quantum-hardware-specific optimizers}, where the gradient evaluation requires some quantum architecture. The availability of multiple options leads to confusion in the choice of proper optimizers for measurement-specific or otherwise overall performance. In that respect, a benchmarking of the optimizers specific to quantum chemistry applications is essential. Similar benchmarking studies are also available in the literature  on variational quantum linear solvers~\cite{linear} and quantum machine learning (QML) problems~\cite{Joshi_2021}. There exists one benchmarking study on quantum chemistry problems~\cite{chem2}, where only a few specific gradient-free optimizers were considered. \\

\noindent The aim of the present work is to  provide a comparative performance analysis of some of the commonly used classical optimizers in the variational quantum algorithms based on different applications related to quantum chemistry.
The comparison has been made with two different simulators, the ideal quantum circuit simulator (Qiskit's statevector\_simulator) and the noisy quantum circuit simulator (Qiskit's qasm\_simulator). The latter provides us with an idea of the performance of these classical optimizers in ideal and noisy conditions.
To that end, five molecular systems (H$_2$, LiH, BeH$_2$, H$_2$O, HF) are considered to form a reasonable set  with a different number of electrons (2 to 10), different natures of chemical bonding (covalent or ionic), and different molecular geometries (linear or bent). Quantum simulations of these molecules were achieved using optimizers of different classes, such as gradient-based, gradient-free, and quantum-hardware-specific optimizers. The efficiency of these optimizers is benchmarked against  convergence criteria of simulations and accuracies of molecular properties, such as total energies, dissociation energies, and dipole moments. 

\begin{figure*}[!]
\includegraphics[width=15cm]{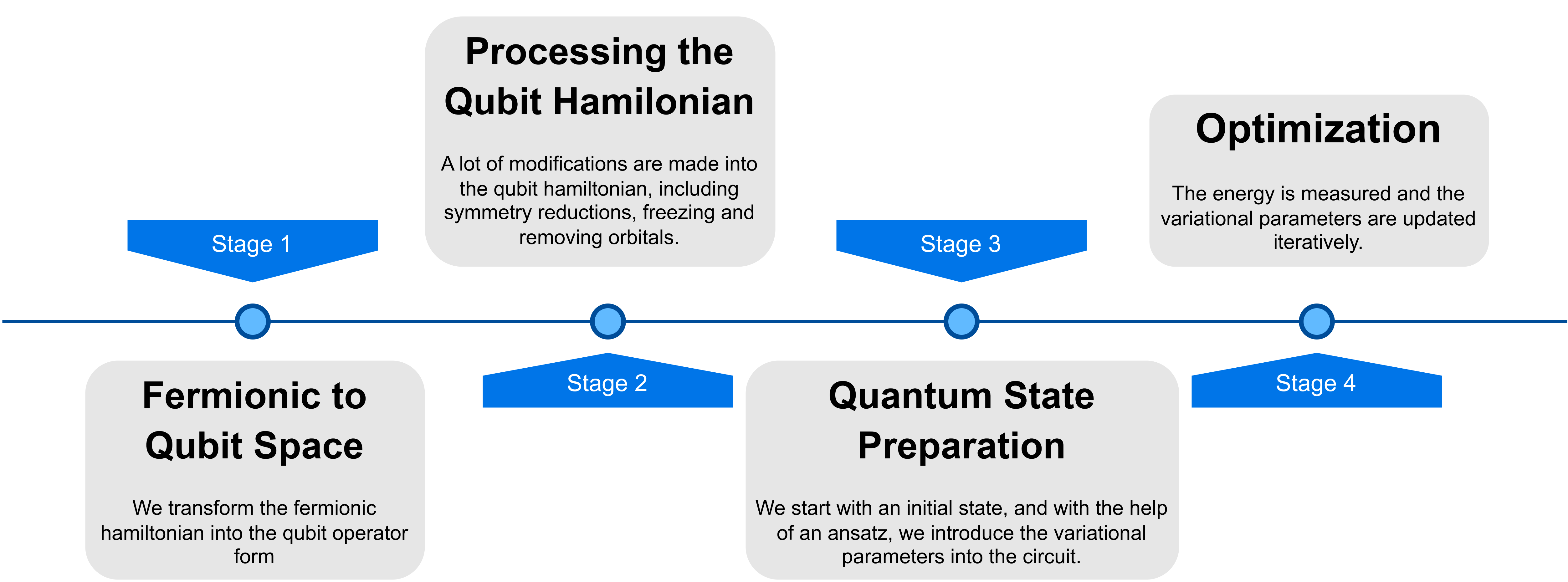}
\caption{\label{fig:stages}Workflow diagram highlighting the application of a variational quantum algorithm in quantum chemistry.}
\end{figure*}

\section{Chemistry on a Quantum Computer}

\noindent VQAs have been used extensively in the literature for applications in quantum chemistry~\cite{intro1} with diligence on the theoretical development of algorithms and  numerical techniques~\cite{intro2,intro3}, and their implementations~\cite{intro4,intro5}. FIG.~\ref{fig:stages} shows the various steps involved in the application of a variational quantum algorithm in quantum chemistry.\\

\noindent The VQAs are based on the variational principle, which states that the ground state energy of the system is the lower bound for its energy spectrum. Hence any arbitrary state will have energy more than or equal to the ground state energy. In a variational problem, the state space of a given setup is explored to find the state that corresponds to the minimum energy. For a molecular system with $N$ electrons, and $M$ nuclei with nuclear charge $Z$ and nuclear mass $M$, the Hamiltonian (in atomic unit) is given by,
\begin{eqnarray}
H=&& - \sum_{A=1}^{M} \frac{\nabla_A^2 }{2M_A} -\sum_{i=1}^{N} \left( \frac{\nabla_i^2}{2}  + \sum_{A=1}^{M} \frac{Z_A}{r_{iA}} \right)\nonumber\\
&&+  \sum_{j>i}^N\frac{1}{r_{ij}} + \sum_{B>A}^M\frac{Z_A Z_B}{R_{AB}}.
\label{eq:2}
\end{eqnarray}
Under the Born-Oppenheimer approximation, the Hamiltonian is further simplified for the electrons to the form,
\begin{equation}
H=-\sum_{i=1}^{N} \left( \frac{\nabla_i^2}{2}  - \sum_{A=1}^{M} \frac{Z_A}{r_{iA}} \right) + \sum_{j>i}^{N}\frac{1}{r_{ij}}.
\label{eq:3}
\end{equation}

\noindent Quantum computation of the quantum chemistry problems defined by the molecular Hamiltonian operator requires a reformulation in the fermionic space in terms of the $2^N$-dimensional qubit space. This conversion is easily achieved in the Fock space representation, where the wave function can be written in terms of occupation numbers,
\begin{equation}
    |\vec k\rangle = |k_1,k_2, \cdots ,k_N\rangle,\quad k_p= \{0, 1\}.
    \label{eq:4}
\end{equation}
Here, $k_p = 0$ and $1$ signify the $p^{th}$ (spin)-orbital as unoccupied and occupied, respectively. In this representation of the wavefunction, the one-to-one correspondence between the fermionic and the qubit space is straightforward, i.e.,
\begin{equation}
    |\vec k\rangle = |k_1,k_2, \cdots ,k_N\rangle \rightarrow |\vec q\rangle = |q_1,q_2, \cdots ,q_N\rangle
    \label{eq:5}
\end{equation}
where, each orbital and its occupancy ($k_p= \{0,1\}$) is represented by the state of a qubit $q_p= \{\uparrow , \downarrow\}$.
The operators in the Fock-space representation are expressed in terms of the creation and annihilation operators~\cite{szabo}, ($a^\dagger_p$ and $a_p$, respectively), defined by:
\begin{eqnarray}
    a^\dagger_p|\vec k\rangle &=& (1-\delta_{k_p,1}) \Gamma^k_p |k_1,k_2, . ,1_p, \cdots, k_n\rangle \label{eq:6} \\
    a_p|\vec k\rangle &=& \delta_{k_p,1} \Gamma^k_p |k_1,k_2, . ,0_p,\cdots ,k_n\rangle. \label{eq:7}
\end{eqnarray}
\noindent where
\begin{equation}
  \Gamma^k_p = (-1)^{ \sum_{m<p}k_m}.
\end{equation}
\noindent In the second quantization, the molecular Hamiltonian is expressed as,
\begin{equation}
    H = \sum_{p,q} h_{pq} a_p^\dagger a_q + \frac{1}{2}\sum_{p,q,r,s}h_{pqrs}a_p^\dagger a_q^\dagger a_s a_r 
    \label{eq:8}
\end{equation}
where, $h_{pq}$ and $h_{pqrs}$ are the one-electron and the two-electron integrals, providing the electron-nucleus and electron-electron interactions, respectively. They can be defined with the help of some basis functions $\{X(\Vec{x})\}$ as,
\begin{equation}
    \displaystyle h_{pq} = \int d \vec{x} X_p^*(\vec{x})\left(-\frac{\nabla^2}{2}-\sum_{A}\frac{Z_{A}}{r_{A \vec x}}\right)X_q(\vec{x})
    \label{eq:9}
\end{equation}
and 
\begin{equation}
    h_{pqrs} = \iint d \vec{x}_1d \vec{x}_2\frac {X_p^*(\vec{x}_1)X_q^*(\vec{x}_2)X_r(\vec{x}_1)X_s(\vec{x}_2)}{r_{12}}.
\end{equation}

\noindent The operators in the Fock representation can be transformed to the qubit-space by various transformation schemes, such as the Jordan-Wigner~\cite{jw}, Parity~\cite{parity}, and Brayvi-Kitaev~\cite{bk} schemes. In Jordan-Wigner representation, the transformation rules are,
\begin{eqnarray}
    a_i^\dagger  &=& \frac{1}{2} (X_i - iY_i) \otimes_{j<i} Z_j \label{eq:tr_rule1}\\
    a_i &=& \frac{1}{2} (X_i + iY_i) \otimes_{j<i} Z_j \label{eq:tr_rule2},
\end{eqnarray}
\noindent where, $a_i$ and $a_i^\dagger$ are the ladder operators defined in the fermionic space, and $X, Y, Z$ are the Pauli operators defined in the qubit space. With these transformation rules, the Hamiltonian in the  fermionic space (Eq.~\ref{eq:3} or Eq.~\ref{eq:8}) can be written in the qubit space. The number of gates and qubits in the qubit Hamiltonian would depend on the system of interest. For example, in the case of the H$_2$ molecule, the qubit Hamiltonian can be written as \cite{HydrogenQubit},

\begin{equation}
    H_{\rm qubit}^{\rm H_2} = (-0.81261) IIII + \cdots + (0.17434) ZZII.
\end{equation}
\noindent With the given qubit Hamiltonian for a molecular system, the quantum measurement of its energy is achieved by following the strategy described in FIG~\ref{fig:vqa}. 


Starting from the Hartree-Fock state as the initial state $ |\Psi_{\rm in}\rangle$, within the coupled cluster (CC) ansatz the excitation operator is expressed as a sum of 
clusters of excitations, i.e., $T = \sum_{i}^{n} T_{i}$, with $T_i$ representing  $i$-electron excitations from $N$ electrons. 
Generally, in the CC method, an exponential ansatz is used, which is truncated at some fixed level of excitations. When truncation at the second-excitation level is chosen (the so-called CC singles and doubles (CCSD) method), the wavefunction is given by,
\begin{equation}
    |\Psi({\rm CCSD})\rangle = e^{T_1 + T_2} |\Psi_{\rm in}\rangle,
\end{equation}
with 
\begin{equation}
    T_1 = \sum_{\substack{i\in \rm occ \\ k\in \rm virt}} t_k^i a^{\dagger}_k a_i
\end{equation}
and 
\begin{equation}
    T_2 = \sum_{\substack{i>j\in \rm occ \\ k>l\in \rm virt}} t_{kl}^{ij} a^{\dagger}_k a^{\dagger}_l a_i a_j
\end{equation}
representing the single- and double-excitation operators, respectively, from the occupied orbitals ($i,j$) to the virtual orbitals $(k,l)$. The expansion coefficients
$t_k^i$, and $t_{kl}^{ij}$ account for the contribution of the corresponding excitation.
For applications of the CC ansatz on a quantum computer, we  redefine the excitation operator as  unitary; hence the name unitary CC (UCC)~\cite{ucc}, 
\begin{equation}
    |\Psi({\rm UCC})\rangle = e^{T - T^{\dagger}} |\Psi_{\rm in}\rangle.
\end{equation}

\noindent We can write the excitation operators in terms of qubit operators using any standard mapping techniques like the Jordan-Wigner method truncated at a particular excitation level and then use in a variational quantum circuit~\cite{uccc}.
    
\begin{figure*}[!]
\includegraphics[width=15cm]{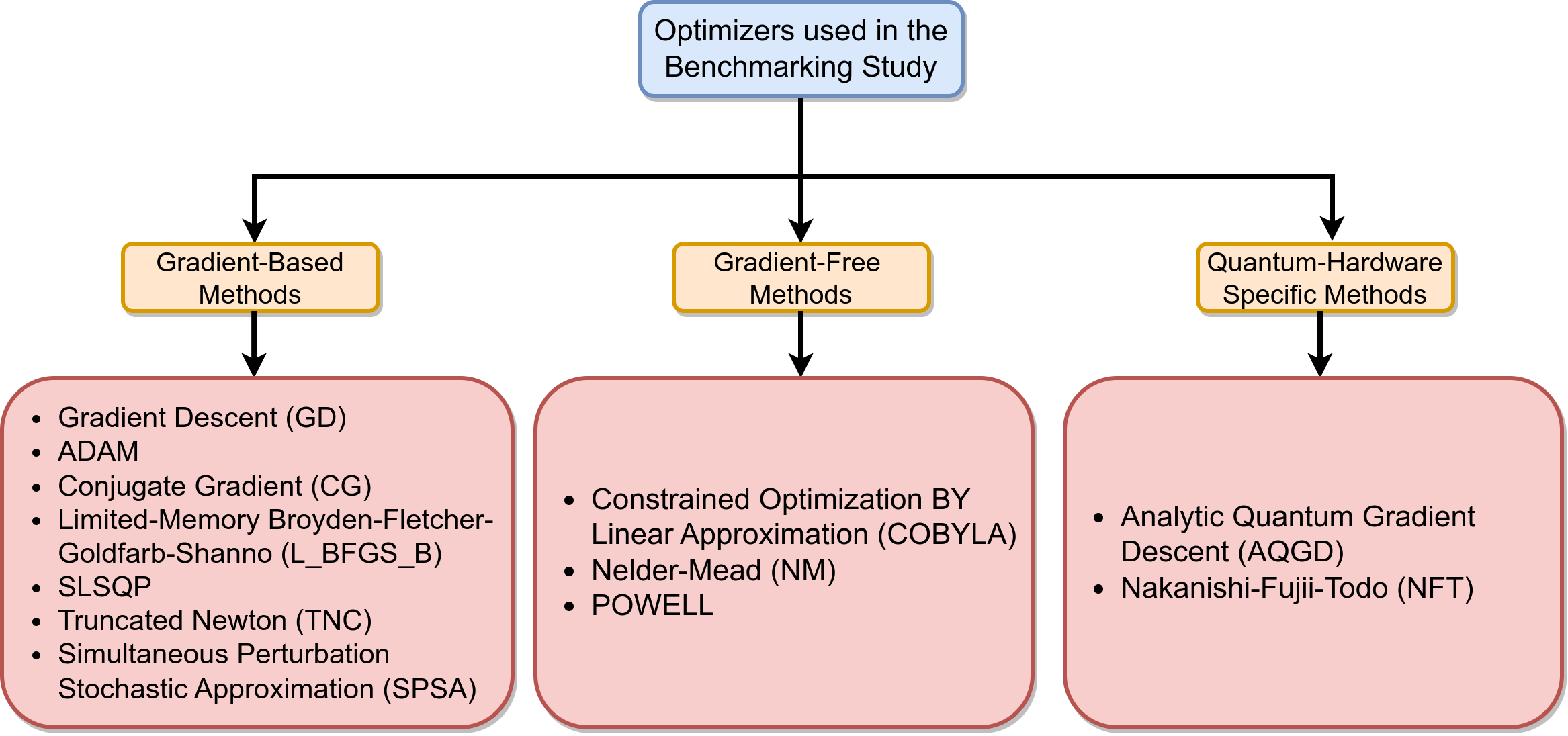}
\caption{\label{fig:flowchart}Different optimizers used in this benchmarking study, which are broadly classified into three categories: (a) Gradient-based methods, (b) Gradient-free methods, and (c) Quantum-Hardware specific methods.}
\end{figure*}

\section{Optimizers}
\label{sec:Optimizers}
\noindent FIG.~\ref{fig:flowchart} presents the different optimizers that are considered in this benchmarking study. The optimizers are classified into three categories: (a) Gradient-based methods, (b) Gradient-free methods, and (c) Quantum-Hardware specific methods. The simplest way of optimizing a function $f(\theta)$ is to take the help of its gradient, which is defined as $\nabla f(\theta) = \frac{df}{d \theta }$.
The gradient points toward the direction of the greatest change of the function and hence can be used to optimize the objective function. Simple algorithms like the Gradient Descent (GD)~\cite{grad} use the gradient directly for optimization. The parameter update rule in the GD algorithm for minimization is given by, 
\begin{equation}
   \theta_{n+1} = \theta_{n} - h \left. \frac{df}{d\theta}\right|_{\theta = \theta_n}, 
\end{equation}
where $h$ is the so-called learning rate. Analyzing from a given point and the gradient at that point, we predict how far away it can go, or, in other words, `learn'. All the algorithms that require the evaluation of the gradient are hence classified as the gradient-based methods~\cite{adam, cg, cg1, bfgs, lbfgsb, slsqp, tnc}. However, there are many occasions where the objective function is unknown or the evaluation of the gradient is difficult, and in those cases, we require optimizers to work without any information regarding the gradient of the objective function. These optimizers are classified as gradient-free methods~\cite{cobyla, nm, Powell, spsa}. Finally, we have a class of special optimizers that require a quantum component in optimization and are hence placed differently under the quantum-hardware specific methods~\cite{aqgd, nft}.

\section{Methodology}
\noindent Five different molecules (H$_2$, LiH, BeH$_2$, H$_2$O, and HF) having different levels of chemical and numerical intricacies are considered in this work for the benchmarking study (Table~\ref{tab:table1}). For each of these molecular systems, we use two numerical methods for the energy evaluation, i) the variational quantum eigensolver (VQE), which finds the ground state of the Hamiltonian through the variational principle, and ii) the Numpy eigensolver, which diagonalizes the Hamiltonian and provides the classical (numerically exact) results. 
For the quantum simulation, we use the UCC ansatz~\cite{ucc}. Within the UCC ansatz, the requested excitations are constructed  starting from the Hartree-Fock reference state. Both single and double excitations are considered for accurate treatment of electron correlation. The energy is evaluated over a range of geometries for each molecule by varying the bond distances from  0.1 to 4~\AA. For the triatomic systems, the molecule is distorted along the totally symmetric stretching mode. We estimate the equilibrium geometry, the equilibrium energy, and the dissociation energy from the resulting potential-energy curves. We carry out this process for all the optimizers mentioned in FIG.~\ref{fig:flowchart}.\\

\begin{table*}
\caption{\label{tab:table1}The Molecules Considered for Simulations, their Active Space, and the Number of Qubits required.}
\begin{ruledtabular}
    \begin{tabular}{cccccc}

Molecule & Electronic Configuration & Active Space & Number of Qubits\footnote{The number of qubits required, in exception to H$_2$, is reduced by a factor of 2 because parity mapping was used and a Z$_2$-symmetry reduction~\cite{taper} was employed. The molecular orbitals are denoted in the notation of the corresponding molecular symmetry point group.}\\
\hline  
\hline  
         H$_2$&  $1\sigma_g^2$  $1\sigma_u^0$ & $1\sigma_g^2$ $1\sigma_u^0$ & 2\\
         LiH  &  $1\sigma^2$  $2\sigma^2$  $3\sigma^0$ & $1\sigma^2$ $2\sigma^2$ $3\sigma^0$ & 4\\
         BeH$_2$ &$1\sigma_g^2$  $2\sigma_g^2  1\sigma_u^2$  $1\pi_g^0 1\sigma_g^{*0}
         1\sigma_u^{*0}$ &  $2\sigma_g^2  1\sigma_u^2$  $1\pi_g^0$ & 6\\
         H$_2$O & $1a_1^2$  $2a_1^2 1b_2^2 3a_1^2 1b_1^2$   $4a_1^0 2b_2^0$ & $ 1b_2^2 3a_1^2 1b_1^2$   $4a_1^0 2b_2^0$ & 8\\
         HF &  $1\sigma^2$ $2\sigma^2 3\sigma^2 1\pi^4$   $4\sigma^{*0}$ &  $1\sigma^2 2\sigma^2 3\sigma^2 1\pi^4$ $4\sigma^{*0}$ & 10
\label{table4}
\end{tabular}
\end{ruledtabular}
\end{table*}

\noindent The performance of the employed optimizers in VQE for molecular simulations is  assessed by evaluating the ground state energy error ($\Delta_{\rm gs}$), the dissociation energy error ($\Delta_{\rm de}$), and the root-mean-squared dipole-moment error ($\displaystyle \Delta^{\rm MSE}_{\rm dipole}$), given by the following expressions.
\begin{equation}
    \Delta_{\rm gs} = \frac{E_{\rm gs}^{\rm exact} - E_{\rm gs}^{\rm VQE}}{E_{\rm gs}^{\rm exact}}, \label{eq:gs_err}
\end{equation}
where $E_{\rm gs}^{\rm exact}$ and $E_{\rm gs}^{\rm VQE}$ are the ground state energy as evaluated by the Numpy eigensolver and by the VQE, respectively. 
\begin{equation}
    \Delta_{\rm de} = \frac{E_{\rm de}^{\rm exact} - E_{\rm de}^{\rm VQE}}{E_{\rm de}^{\rm exact}},\label{eq:de_err}
\end{equation}
where, $E_{\rm de} = E_{\infty} - E_{\rm gs}$. $E_{\infty}$ is approximated as the energy of the molecule with an inter-atomic distance of 4~\AA. 
\begin{equation}
\displaystyle \Delta^{\rm MSE}_{\rm dipole} = \sqrt{\frac{ \displaystyle \sum^{N}_{i} \left( \mu_{i}^{\rm exact} - \mu_{i}^{\rm VQE} \right)^{2}}{N}}, \label{eq:dm_err} \end{equation} 
where $\mu_i$ is the dipole moment of the non-centrosymmetric molecule at one of the $N$ geometries. The mean-squared dipole moments require calculations of dipole-moment at different inter-nuclear distances, which is useful for studying vibronic spectroscopy and accurately describing chemical bonding. However, the dipole moment at different inter-atomic distances is not uniformly linear, see the work for LiH~\cite{lihdipole}. The non-linear dependence of the dipole moment on inter-atomic distance is observed at a small inter-nuclear distance (where Pauli repulsion dominates) and at a large inter-nuclear distance (where long-range interactions are present). The dipole moment is linear with the inter-nuclear distance at the intermediate region For this reason, here, the dipole moments for LiH, H$_2$O, and HF  are evaluated over the inter-nuclear distances of $1.3-2.0$~\AA,  $0.7-1.2$~\AA,  and $0.8-1.4$~\AA, respectively. 

The maximum number of iterations was kept to $100$ for all optimizers, which ensures that all the optimizers use the same computational resource for the task, hence making the comparison fair. To add another level of scrutiny, we then assessed the best-performing optimizers in these tasks with the minimum number of iterations required to achieve a desired tolerance level (in this case, the tolerance was chosen to be $10^{-6}$). This further helps to decide the overall performance of an optimizer. Since the currently available quantum hardware is noisy~\cite{nisq}, an optimizer that requires a larger number of iterations for convergence accumulates more intricate errors as it executes the circuit a larger number of times. Hence, the optimizers that converge in a smaller number of iterations are desirable with noisy hardware. 


All the simulations were first performed with an ideal quantum circuit simulator (statevector\_simulator provided by Qiskit). This gives us an idea about which optimizers should work well with the quantum hardware in a perfect situation. However, with the current state of noisy intermediate-scale quantum (NISQ) devices, the effect of noise should also be considered. To effectively understand the impact of the noise on the performance of the classical optimizers, two additional sets of simulations were carried out. First, the benchmarking is done with a noisy quantum circuit simulator (qasm\_simulator provided by Qiskit), which only has the shot-noise. The next set of simulations was performed with a noisy quantum circuit simulator (qasm\_simulator) with additional noise embedded from a noise model sampled from the IBM Cairo quantum device. The latter set of simulations is, therefore, equivalent to the performance of a realistic fault-tolerant quantum computer. This strategy has been followed by the previous benchmarking works for the variational quantum linear solvers~\cite{linear}.

\section{Results and Discussion}

\begin{figure*}
\includegraphics[width=18cm]{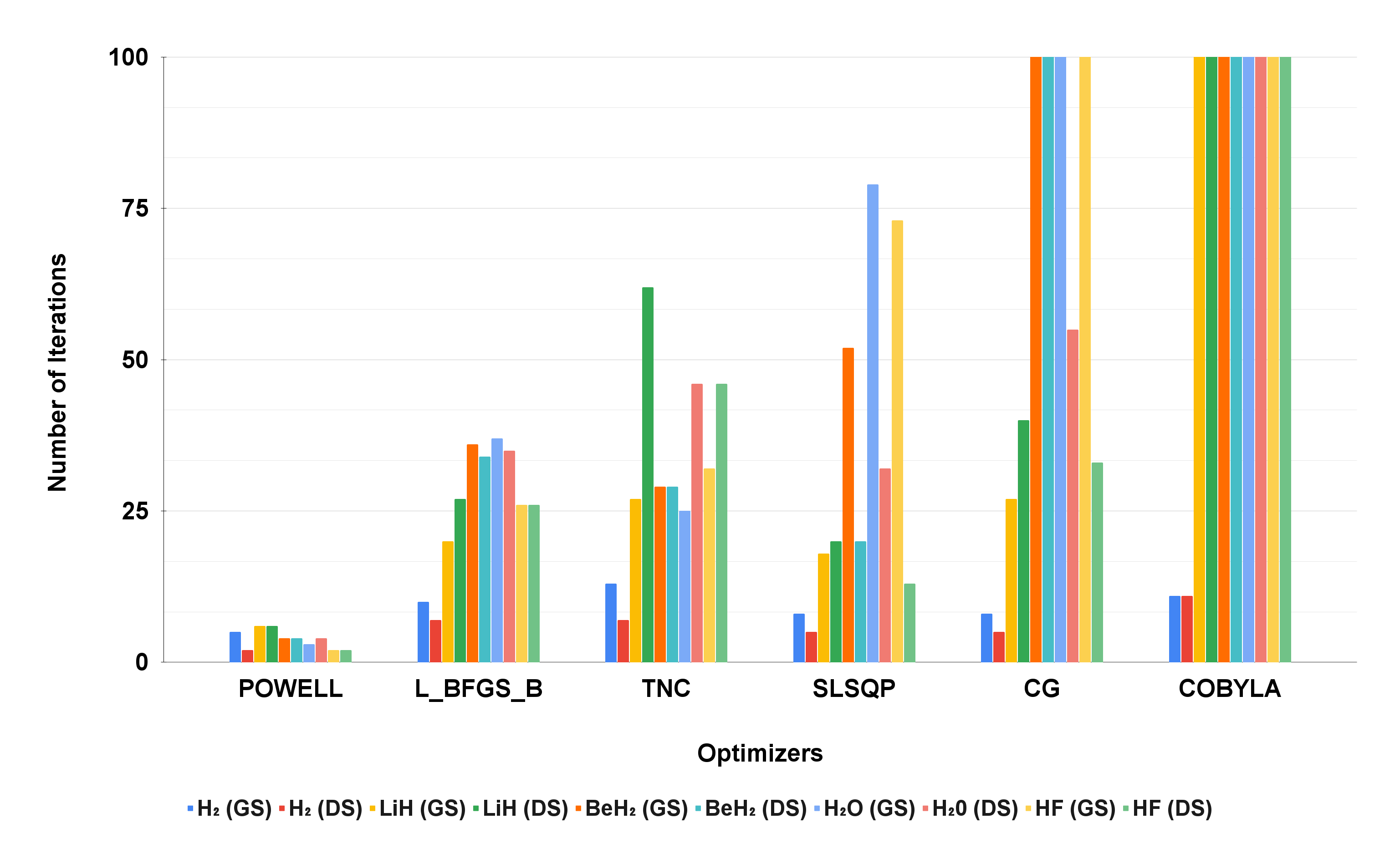}
\caption{\label{fig:iter}Number of iterations taken by different optimizers for convergence (Simulations run on an ideal quantum circuit simulator). The chosen tolerance is 10$^{-6}$, and for each molecule, two simulations are done, the ground state energy error (GS) and dissociation energy error (DS). The Optimizers CG and COBYLA take more than 100 iterations to converge.}
\end{figure*}

\noindent TABLE~\ref{tab:table2} presents the errors (i.e., the difference between the VQE and the Numpy results) of the ground state energy, the dissociation energy, and the mean-squared dipole-moment and the average error in these properties across all the considered molecular systems, obtained from quantum simulation based on different optimizers done with the ideal quantum circuit simulator. In this case, the maximum number of iterations for each optimizer is set to 100, which allows for a fair assessment of the performance of these optimizers. Several optimizers, namely CG, SLSQP, POWELL, and COBYLA perform remarkably well. The results for the convergence speed of the optimizers that performed well can be seen in FIG.~\ref{fig:iter}. For the chosen tolerance of $10^{-6}$, two properties, namely the ground state energy and dissociation energy error, for each molecule have been observed. The results show a stark difference in performance, with POWELL being clearly the fastest converging optimizer and even holding up its performance for larger molecules. L\_BFGS\_S, TNC, and SLSQP also perform well, each achieving the tolerance level within 100 iterations. On the other hand, the SPSA, AQGD, and NFT optimizers perform poorly, \textit{vide infra}. 

\noindent Table~\ref{tab:table3} presents the ground state energy, the dissociation energy, and the mean-squared dipole-moment and the average error in these properties for four molecular systems (H$_2$, LiH, BeH$_2$, and H$_2$O), obtained from quantum simulation based on different optimizers done with the noisy quantum circuit simulator. HF molecule could not be considered in this set due to time and computational constraints, as the noisy simulators are often very slow when compared to the ideal simulators, especially for large qubit systems. Table~\ref{tab:table4} presents the simulation results obtained from circuits run with the noisy quantum circuit simulator and embedded noise from the noise model sampled from the IBM Cairo quantum computer. While the performance of all the optimizers was further hampered due to the addition of more noise into the circuit, the performance trend from Table~\ref{tab:table3} remains largely unchanged, with SPSA, COBYLA, POWELL, and GD being the best-performing optimizers even in the presence of noise.



\begin{table*}
\caption{\label{tab:table2}Performance of the classical optimizers with an ideal quantum circuit simulator across different parameters, namely the ground state energy error (Eq.~\ref{eq:gs_err}, error in dissociation energy (Eq.~\ref{eq:de_err}), and dipole moment-mean squared error  (Eq.~\ref{eq:dm_err}) for different molecules, where the maximum number of iterations is set to 100.}
\begin{ruledtabular}
\begin{tabular}{ccccc}
Optimizer & $\Delta_{\rm gs}$ & $\Delta_{\rm de}$ & $\Delta^{MSE}_{\rm dipole}$  & Average Error\\
\hline
\hline

CG & $8.76\times 10^{-9}$ & $1.78\times 10^{-5}$& $1.26\times 10^{-5}$& $1.02\times 10^{-5}$\\
\hline
SLSQP & $1.53\times 10^{-8}$ & $1.69\times 10^{-5}$& $6.04\times 10^{-4}$& $2.06\times 10^{-4}$\\
\hline
POWELL & $5.32\times 10^{-7}$ & $2.91\times 10^{-2}$& $1.22\times 10^{-2}$& $1.38\times 10^{-2}$\\
\hline
COBYLA & $1.18\times 10^{-4}$ & $1.42\times 10^{-1}$& $5.74\times 10^{-2}$& $6.68\times 10^{-2}$\\
\hline
L\_BFGS\_B & $8.76\times 10^{-9}$ & $3.49\times 10^{-1}$& $5.40\times 10^{-5}$& $1.16\times 10^{-1}$\\
\hline
TNC & $1.49\times 10^{-6}$ & $7.02\times 10^{-1}$& $2.32\times 10^{-2}$& $2.42\times 10^{-1}$\\
\hline
SPSA & $5.06\times 10^{-3}$ & $9.07\times 10^{-1}$& $6.57\times 10^{-1}$& $5.23\times 10^{-1}$\\
\hline
GD & $6.02\times 10^{-5}$ & $1.7708$& $2.15\times 10^{-2}$& $5.97\times 10^{-1}$\\
\hline
ADAM & $1.83\times 10^{-4}$ & $2.2472$& $2.54\times 10^{-2}$& $7.57\times 10^{-1}$\\
\hline
NM & $1.81\times 10^{-4}$ & $2.3430$& $4.70\times 10^{-2}$& $7.96\times 10^{-1}$\\
\hline
AQGD & $3.50\times 10^{-3}$ & $2.7535$& $6.91\times 10^{-2}$& $9.42\times 10^{-1}$\\
\hline
NFT & $3.50\times 10^{-3}$ & $2.7535$& $6.91\times 10^{-2}$& $9.42\times 10^{-1}$\\

\end{tabular}
\end{ruledtabular}
\end{table*}

\begin{table*}
\caption{\label{tab:table3}Performance of the classical optimizers with a noisy quantum circuit simulator (QASM simulator) across different parameters, namely the ground state energy error (Eq.~\ref{eq:gs_err}, error in dissociation energy (Eq.~\ref{eq:de_err}), and dipole moment-mean squared error  (Eq.~\ref{eq:dm_err}) for different molecules, where the maximum number of iterations is set to 100.} 
\begin{ruledtabular}
\begin{tabular}{ccccc}
Optimizer & $\Delta_{\rm gs}$ & $\Delta_{\rm de}$ & $\Delta^{MSE}_{\rm dipole}$  & Average Error\\
\hline
\hline
POWELL & $ 2.22 \times 10^{-3} $ & $1.12 \times 10^{-1}$& $4.23 \times 10^{-1}$& $ 1.79 \times 10^{-1} $\\
\hline
SPSA & $2.97 \times 10^{-3}$ & $1.68 \times 10^{-1} $& $4.11 \times 10^{-1}$& $ 1.94\times 10^{-1}$\\
\hline
COBYLA & $3.43\times 10^{-3} $ & $6.14\times 10^{-1} $& $1.35\times 10^{-1} $& $ 2.51\times 10^{-1}$\\
\hline
GD & $3.92 \times 10^{-3}$ & $4.43 \times 10^{-1}$& $5.65 \times 10^{-1} $& $3.37 \times 10^{-1}$\\
\hline
NFT & $4.82 \times 10^{-3}$ & $1.2254 $& $1.67 \times 10^{-1} $& $ 4.65 \times 10^{-1}$\\
\hline
ADAM & $5.02 \times 10^{-3}$ & $1.4468 $& $ 6.58\times 10^{-2}$& $ 5.05\times 10^{-1}$\\
\hline
SLSQP & $2.54 \times 10^{-1}$ & $1.0847 $& $9.02 \times 10^{-1}$& $ 7.47\times 10^{-1}$\\
\hline
NM & $1.01 \times 10^{-3}$ & $2.2126 $& $ 2.93 \times 10^{-2}$& $ 7.47\times 10^{-1}$\\
\hline
CG & $ 2.89 \times 10^{-3}$ & $2.2299 $& $ 3.08\times 10^{-2}$& $ 7.54\times 10^{-1}$\\
\hline
TNC & $5.21\times 10^{-3}$ & $2.2524$& $3.12\times 10^{-2}$& $7.62\times 10^{-1}$\\
\hline
L\_BFGS\_S & $6.06\times 10^{-3}$ & $2.2548$& $3.61\times 10^{-2}$& $7.65\times 10^{-1}$\\
\hline
AQGD & $1.05\times 10^{-1}$ & $1.2410$& $1.071$& $8.05\times 10^{-1}$\\

\end{tabular}
\end{ruledtabular}
\end{table*}

\begin{table*}
\caption{\label{tab:table4}Performance of the classical optimizers with a noisy quantum circuit simulator (QASM simulator) and noise embedded from IBM Cairo quantum machine across different parameters, namely the ground state energy error (Eq.~\ref{eq:gs_err}, error in dissociation energy (Eq.~\ref{eq:de_err}), and dipole moment-mean squared error  (Eq.~\ref{eq:dm_err}) for different molecules, where the maximum number of iterations is set at 100.} 
\begin{ruledtabular}
\begin{tabular}{ccccc}
Optimizer & $\Delta_{\rm gs}$ & $\Delta_{\rm de}$ & $\Delta^{MSE}_{\rm dipole}$  & Average Error\\
\hline
\hline
SPSA & $1.27 \times 10^{-1}$ & $4.43 \times 10^{-1} $& $8.41 \times 10^{-2}$& $ 2.18\times 10^{-1}$\\
\hline
COBYLA & $1.22\times 10^{-1} $ & $5.90\times 10^{-1} $& $1.05\times 10^{-1} $& $ 2.72\times 10^{-1}$\\
\hline
POWELL & $ 1.22 \times 10^{-1} $ & $5.58 \times 10^{-1}$& $2.42 \times 10^{-1}$& $ 3.07 \times 10^{-1} $\\
\hline
GD & $1.24 \times 10^{-1}$ & $4.91 \times 10^{-1}$& $9.98 \times 10^{-1} $& $5.38 \times 10^{-1}$\\
\hline
ADAM & $1.27 \times 10^{-1}$ & $1.6959$& $ 1.82\times 10^{-1}$& $ 6.68\times 10^{-1}$\\
\hline
NFT & $1.29 \times 10^{-1}$ & $1.6262 $& $2.88 \times 10^{-1} $& $ 6.81 \times 10^{-1}$\\
\hline
TNC & $1.27\times 10^{-1}$ & $2.3794$& $1.05\times 10^{-1}$& $8.71\times 10^{-1}$\\
\hline
CG & $ 1.27 \times 10^{-1}$ & $2.5673 $& $ 1.10\times 10^{-1}$& $ 9.35\times 10^{-1}$\\
\hline
L\_BFGS\_S & $1.27\times 10^{-1}$ & $2.7139$& $1.14\times 10^{-2}$& $9.81\times 10^{-1}$\\
\hline
NM & $1.27 \times 10^{-1}$ & $2.7136 $& $ 1.14 \times 10^{-1}$& $ 9.85\times 10^{-1}$\\
\hline
SLSQP & $1.27 \times 10^{-1}$ & $1.7139$& $ 1.3577 $ & $ 1.0665$\\
\hline
AQGD & $2.94\times 10^{-1}$ & $1.6179$& $2.3029$& $1.4050$\\

\end{tabular}
\end{ruledtabular}
\end{table*}

\subsection{\label{sec:spsa}Performance of the SPSA Optimizer in Ideal Conditions}

\begin{figure}[!]
\includegraphics[width = 9.0cm]{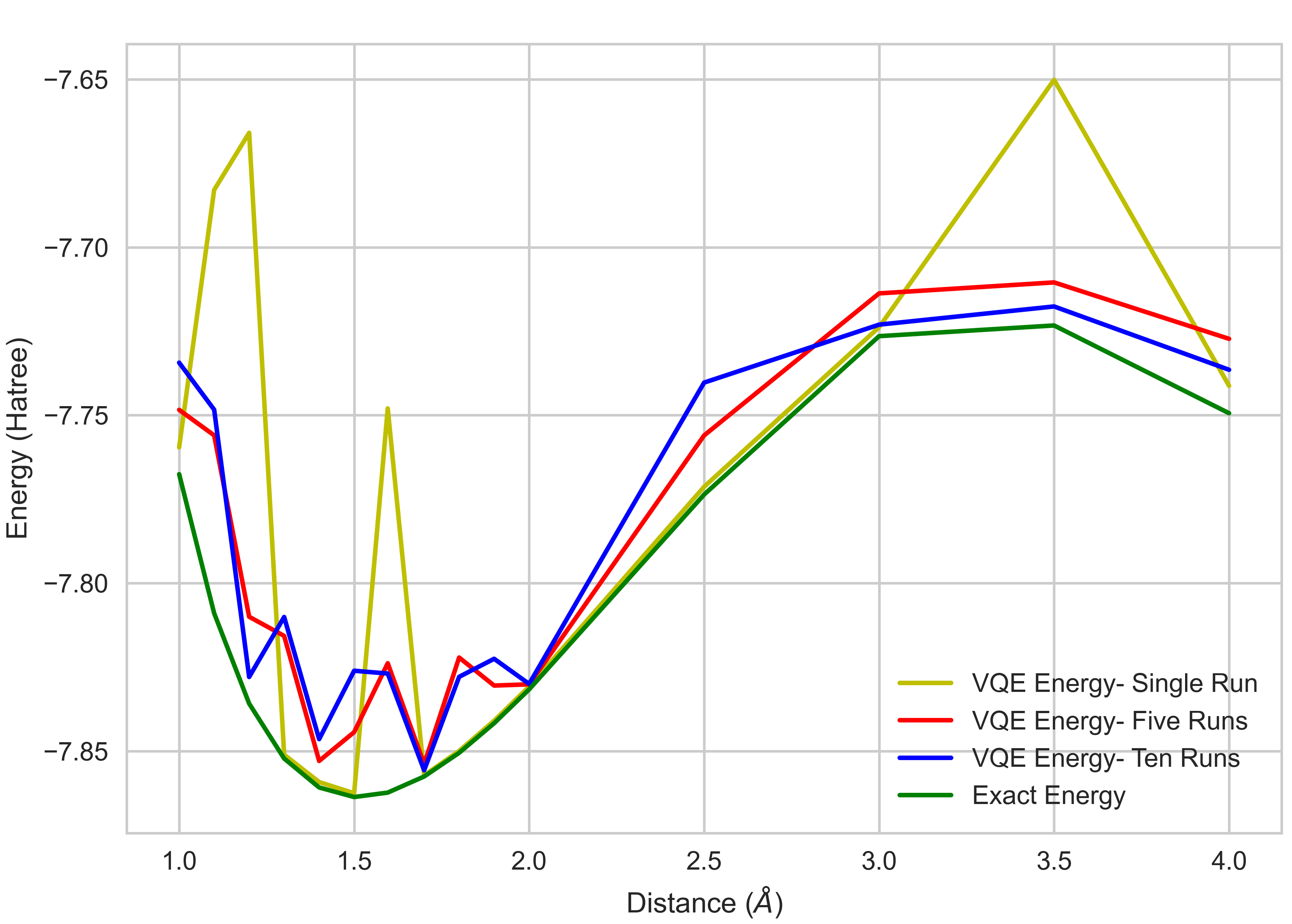}
\caption{\label{fig:spsaavg}Energy Profile of LiH obtained with SPSA optimizer with an ideal quantum circuit simulator and maximum iterations set to 100, averaged over multiple runs.}
\end{figure}

\begin{figure}[!]
\includegraphics[width = 9.0cm]{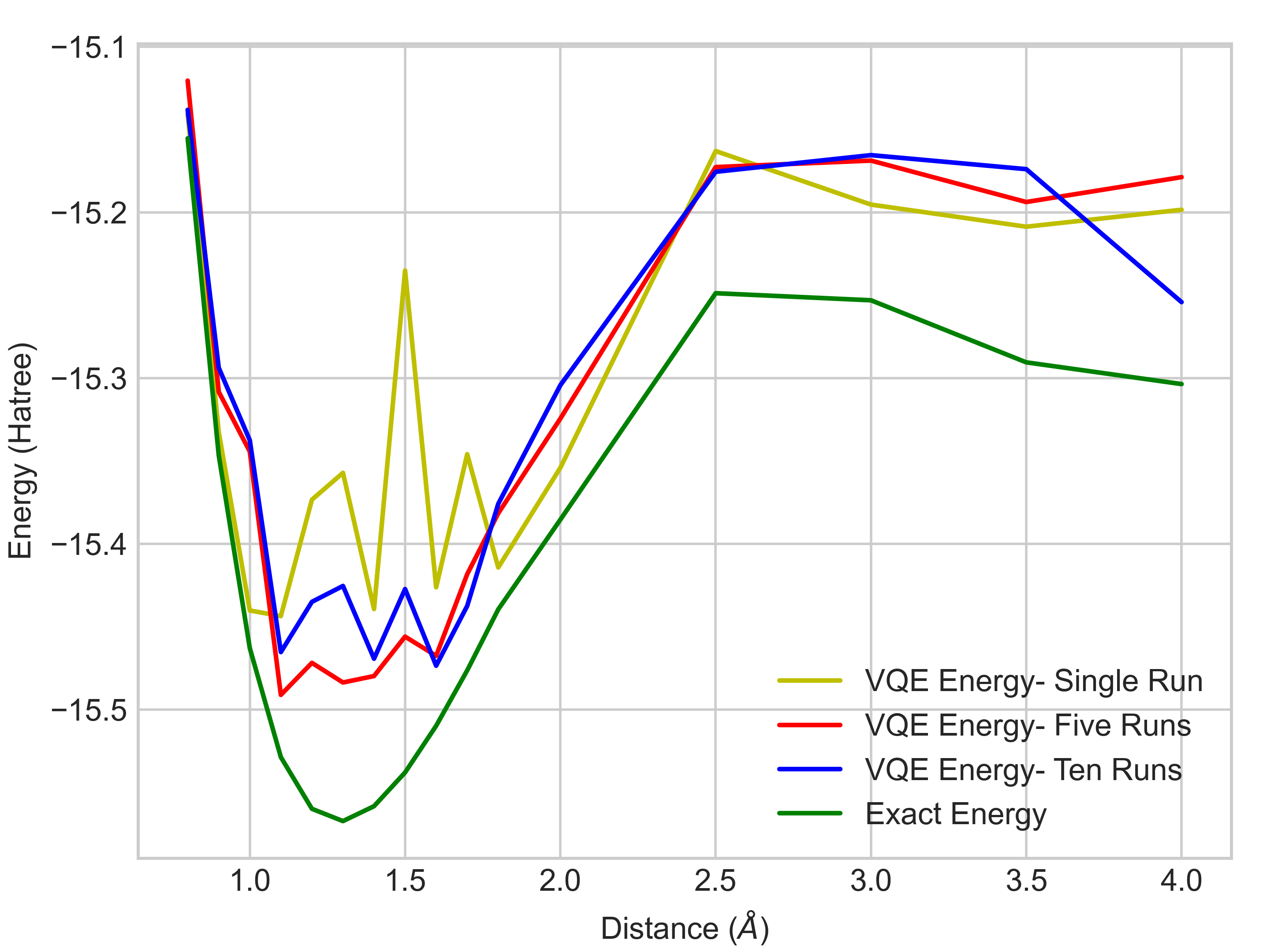}
\caption{\label{fig:spsaavgbe}Energy Profile of BeH$_2$ obtained with SPSA optimizer with an ideal quantum circuit simulator and maximum iterations set to 100, averaged over multiple runs.}
\end{figure}

\noindent  Data from TABLE~\ref{tab:table2} shows that SPSA performs rather underwhelmingly under the ideal simulator conditions, with significant deviations from the exact results, particularly the dissociation energy. The present data can not yield the proper justification behind the large errors in the hybrid classical-quantum algorithm coupled with the SPSA optimizer, a perturbation method where the perturbation is arbitrary and random in nature. As a result, one might need to run the same circuit multiple times to average out these errors, making the algorithm much more computationally expensive. This can be seen in FIG.~\ref{fig:spsaavg}, and FIG.~\ref{fig:spsaavgbe}, where the energy profile of LiH and BeH$_2$ are obtained with SPSA in an ideal condition. It can be seen that by averaging over multiple runs, we get closer to the exact potential-energy curve, and the gap between the VQE results and exact results at larger interatomic distances becomes smaller. Future investigations with a wider range of problems can shed further light on its performance. 

\subsection{Performance of the AQGD optimizer}
\noindent Even though Analytical Quantum Gradient Descent (AQGD)~\cite{aqgd} is one of the modern optimizers, where the gradient is calculated analytically on the quantum hardware, it yields large error for all the benchmarking calculations of the simple simulation of  diatomic molecules with the UCC ansatz compare to an ideal quantum circuit simulator. This trend is also true in noisy simulations where AQGD is the worst-performing optimizer.
The reason for this large error lies in the quantum hardware-specific structure of the AQGD optimizer itself.  
AQGD analytically calculates gradients of the cost function when the circuit consists of a specific set of parameterized gates, namely, gates that have two eigenvalues and can be expressed as
\begin{equation}
   \mathcal{G}(\mu) = e^{-i\mu G} \label{eq:aqgd},
\end{equation}
generated by a Hermitian operator $G$. If we can express a parameterized gate as an ansatz in this way, the gradient of a cost-function $E$, for an operator $ H $, can then be calculated by the `parameter-shift rule' with the shift $ s $ and eigenvalue $r$,
\begin{eqnarray}
\partial_\theta E&=&r \left( \left\langle \psi \left| \mathcal{G}^+(\theta + s)\hat{H} \mathcal{G}(\theta + s) \right| \psi \right \rangle\nonumber \right. \\
&& \left. - \left \langle \psi \left| \mathcal{G}^+(\theta - s)\hat{H} \mathcal{G}(\theta - s) \right| \psi \right\rangle \right ) \label{eq:aqgd_phaseshift} \\
&=& r\left( E(\theta+s) - E(\theta-s)\right). 
\end{eqnarray}
It should be noted that for a quantum circuit with multiple gates having different parameters $\theta_i$, the gradient is to be obtained by using the product rule, shifting corresponding parameters in different gates individually,  and the final result is the sum of these individual terms. 

\begin{figure}[!]
    \centering
        \includegraphics[width=9cm]{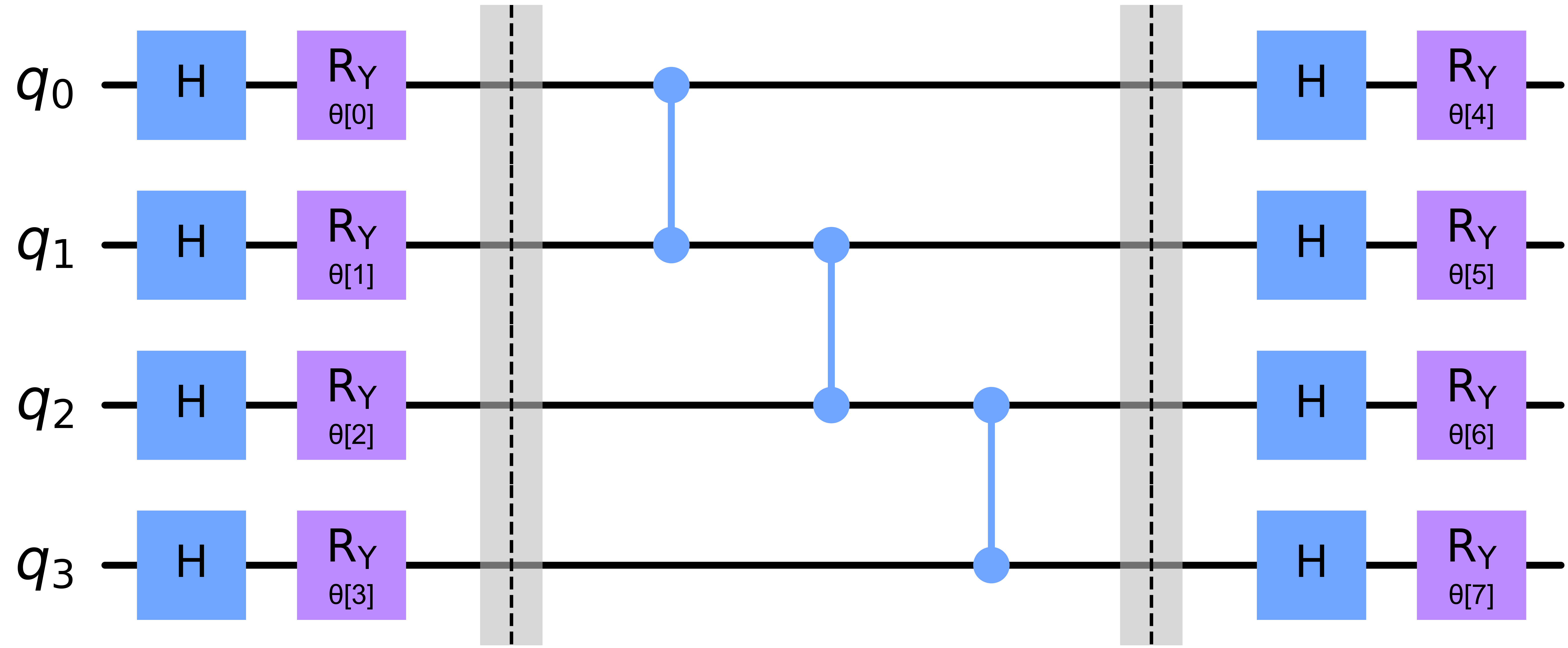}
        \caption{A custom TwoLocal Ansatz with H and R$_y$ as rotation blocks, CZ as the entangling gate, linear entanglement, and a single repetition.}
        \label{fig:twolocal_ansatz}
    \end{figure}

For the particular case where the generator $\mathcal{G}$ is a single-qubit rotation gate, that is, $\mathcal{G} \in \frac{1}{2}\{\sigma_x, \sigma_y, \sigma_z \}$, we have $r = \frac{1}{2}$ and $s = \frac{\pi}{2}$~\cite{aqgd2}.
All the single qubit gates satisfy this condition and hence can be used in an optimization technique involving AQGD. In the UCC ansatz, we consider both single and double excitations that follow a four-parameter shift rule~\cite{anselmetti_2021_local}. This inappropriate adaptation of double excitations explains the large errors observed with the AQGD algorithm coupled with UCC ansatz. To counter that, we prepared a customized ansatz using qiskit's TwoLocal package~\cite{twolocal}, using only single qubit rotation gates (FIG.~\ref{fig:twolocal_ansatz}). The ansatz used in this case consists of the rotation blocks of the Hadamard and $R_y$ rotation gates, full entanglement with CZ entanglement gates, and two repetitions (see FIG.~\ref{fig:twolocal_ansatz}, where an ansatz with similar construction, but with linear entanglement and a single repetition is shown for clarity). This customization results in a significant improvement in the performance of the AQGD optimizer, as seen in TABLE~\ref{tab:table5}, where the ground state and dissociation energy errors for H$_2$ and LiH molecules improve significantly.

\begin{table}
\caption{\label{tab:table5}Performance of the AQGD optimizer for different tasks with UCC and a custom TwoLocal Ansatz. The number of iterations has been fixed to the default value of AQGD, equal to 1000.}
\begin{ruledtabular}
\begin{tabular}{c|ccc}
 Molecule & Property & UCC ansatz & TwoLocal ansatz \\ \hline
 H$_2$ & Ground State &  & \\
 & Energy Error & $1.66\times 10^{-2}$ &$7.68 \times 10^{-12}$\\
 \hline
  & Dissociation && \\
   & Energy Error & $1.47$  &$9.48\times 10^{-6}$\\
   \hline
 LiH & Ground State &  & \\
  & Energy Error & $2.43\times 10^{-3}$ &$1.57\times 10^{-5}$\\
  \hline
  & Dissociation &  & \\
   & Energy Error & $2.02$ &$3.30 \times 10^{-4}$\\
\end{tabular}
\end{ruledtabular}
\end{table}

\subsection{Performance of the NFT optimizer}
\noindent The results from the  Nakanishi-Fujii-Todo (NFT) optimizer~\cite{nft} suffer from the same problem as experienced with the AQGD optimizer. NFT is one of the modern optimization methods designed specifically for hybrid quantum-classical algorithms. It is a sequential optimization method, which is robust against statistical error and is also hyper-parameters free, and has less dependence on the initial parameters. NFT is also gradient-based but requires quantum architecture, similar to AQGD.\\
\noindent The NFT optimization has the following preconditions, (i) the parameters in the variational quantum circuit are independently defined; (ii) the circuit only contains either fixed unitary gates or the single-qubit rotation gates; and (iii) the cost function is the weighted sum of expectation values of individual terms in a Hamiltonian, that is, the cost function can be written as
\begin{equation}
    E(\theta) = \sum_{k=1}^{N} w_k  \left \langle \phi_k \left| U^{\dagger}(\theta)H_k U(\theta)\right| \phi_k \right\rangle
\end{equation}
\noindent where, $ H_k $ are the hermitian operators, like the Hamiltonian in case of Chemistry problems, $\{|\phi\rangle \}^{N}_{k=1}$ are the input states, $w_k$ is the weight of the $k^{th}$ term, and $U(\theta)$ is the effective ansatz operation. 
Let $ {\theta}^{(n)}$ be the parameters of the circuit after $ n $ steps and, $ U_{j}^{(n)}(\theta_{j}) $ be the unitary in which parameters are fixed to be $ \theta_{(n)} $ except for the $ j^{th} $ parameter $ \theta_{(j)}. $ Now, if $ L_{j}^{(n)}(\theta_{j}) $ is the cost function at this stage, we can re-write it as
\begin{equation}
    E_{j}^{(n)}(\theta_{j}) = a_{1j}^{(n)} \cos{(\theta_j - a_{2j}^{(n)})} + a_{3j}^{(n)}
\end{equation}
where, $a_{ij}^{(n)}$ are $\theta$-independent constants. The optimization steps are then as follows:
\begin{itemize}
    \item An index is chosen, either randomly or sequentially, from the set of parameters, $ j_n \in \{ 1, 2, 3, \cdots , N  \} $
    \item Using a quantum device, $ E_{j_n}^{(n-1)}(\theta_{j_n}^{n-1})$, $E_{j_n}^{(n-1)}(\theta_{j_n}^{n-1} + \pi /2)$, and $E_{j_n}^{(n-1)}(\theta_{j_n}^{n-1} - \pi/2)$ are evaluated.
    \item From the quantities evaluated above, $ \theta_{j_n} $, which minimizes the cost function $ E_{j_n}^{(n-1)}(\theta_{j_n}) $ is determined.
    \item Update as follows:
    \begin{equation}
       \theta_{j_n}^{(n)} = \operatorname*{argmin}_{\theta_{j_n}} E_{j_n}^{(n-1)}(\theta_{j_n}) 
    \end{equation}
    \begin{equation}
       \theta_{j}^{(n)} = \theta_{j_n}^{(n-1)}, \quad j \neq j_n 
    \end{equation}
    \begin{equation}
       E_{j_{n+1}}^{(n)}(\theta_{j_{n+1}}^{(n)}) = \operatorname*{min}_{\theta_{j_n}} E_{j_n}^{(n-1)}(\theta_{j_n}).
    \end{equation}
\end{itemize}

\noindent Similar to AQGD, the NFT optimization requires the gates used in the circuit to be specifically single-qubit rotations. The UCC ansatz does not fulfill this requirement and like AQGD, we can obtain significant improvements using a simple ansatz with only single-qubit rotations. Therefore, the results are precisely similar for both optimizers. \\

\noindent The results of all the molecules considered here reveal the optimizers AQGD and NFT to perform identically and therefore, the results are not presented here in the tabulated form. This judgment can be attributed to the fact that both optimizers essentially have similar architecture. 
Consider the AQGD optimizer, where the gradient is evaluated with the phase shift rule as shown in Eq.~\ref{eq:aqgd_phaseshift}.
In NFT, the objective function is first written in terms of a sine function. Then the same phase-shift rule is employed here (the phase-shift, in this case, is then $\frac{\pi}{2}$). If $H$ is the loss-function, then using a quantum device, $ H_{j_n}^{(n-1)}(\theta_{j_n}^{n-1}), H_{j_n}^{(n-1)}(\theta_{j_n}^{n-1} + \pi /2), H_{j_n}^{(n-1)}(\theta_{j_n}^{n-1} - \pi/2) $ are evaluated. From these quantities, the optimized parameter $ \theta_{j_n} $ is determined, which minimizes the cost function $ H_{j_n}^{(n-1)}(\theta_{j_n})$. 

\subsection{Effect of Noise on the Performance of the Optimizer}
\begin{figure}[!]
        \includegraphics[width=8.5cm]{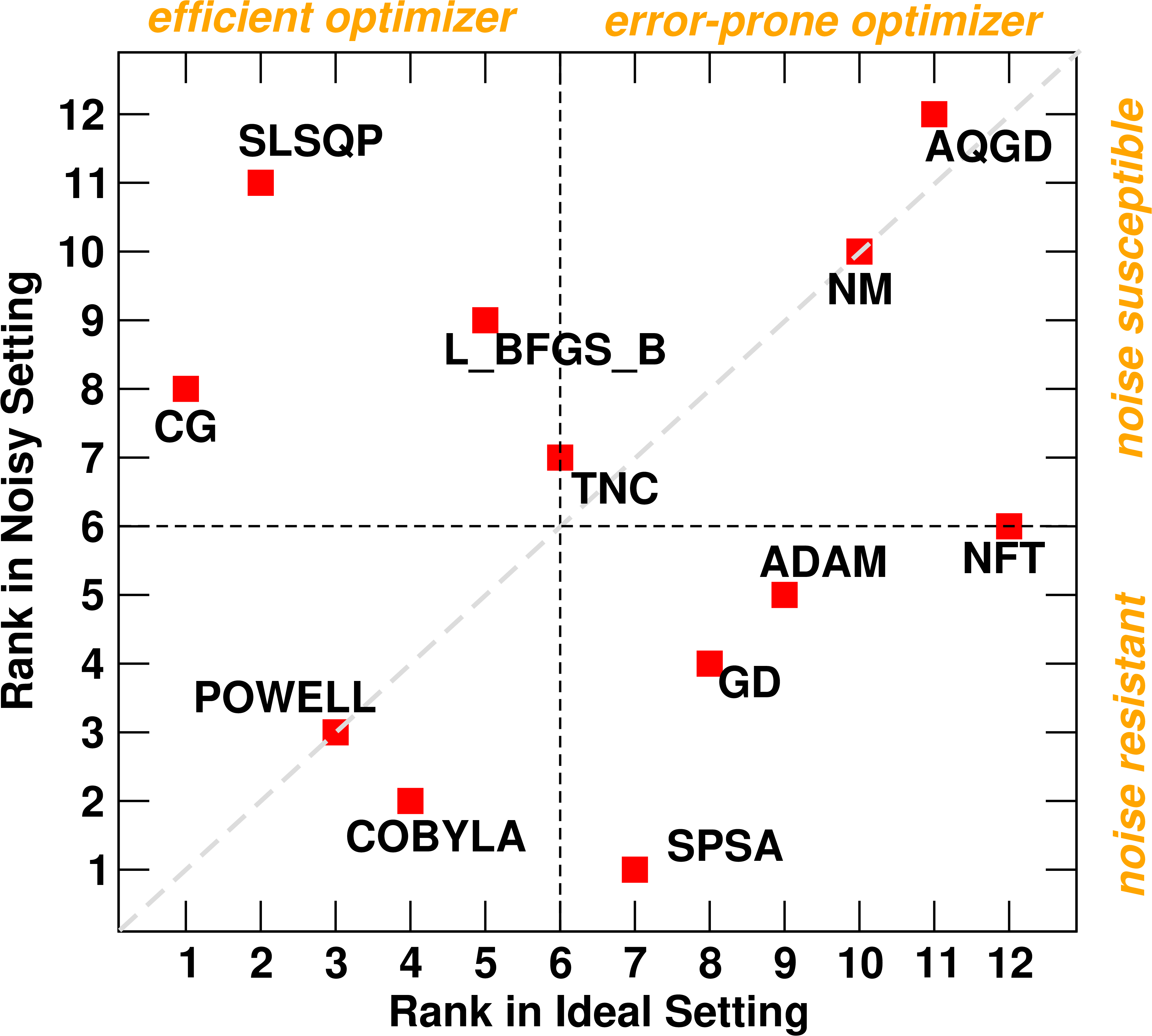}
        \caption{\label{fig:rank}Performance distribution of different classical optimizers in ideal and noisy settings into (i) efficient optimizers, which are among the top performers in an ideal quantum setting, and (ii) error-prone optimizers, which are among the lower-end performers in an ideal setting, (iii) noise-resistant optimizers, returning a decent result even in the presence of noise, and (iv) noise-susceptible, which are the most affected by the noise.}
\end{figure}

\noindent TABLE~\ref{tab:table2} highlights the performance of the classical optimizers in an ideal setting, presenting one with plenty of good options to choose from. However, once the noise is introduced into the circuit, which represents the realistic setting of the modern NISQ machines, many of the optimizers that perform well otherwise suffer a lot. The results in TABLE~\ref{tab:table3} present the optimizer performance when a noisy quantum circuit simulator is employed, which only includes the shot noise, and the results in TABLE~\ref{tab:table4} include results for when additional noise from the IBM Cairo device is embedded into the simulator. The latter is closest to the realistic quantum architecture. Interestingly, the performance trend from TABLE~\ref{tab:table3} to TABLE~\ref{tab:table4} remains somewhat similar, highlighting the fact that the presence of additional noise does not result in drastic changes within the relative performance of the optimizers. The overall results are summarized in FIG.~\ref{fig:rank}, where a visual comparison of optimizers in ideal quantum setting and the noisy quantum setting is presented, based on the results presented in  TABLE~\ref{tab:table2} and TABLE~\ref{tab:table4}, respectively. The performance of the optimizers in the ideal setting can be divided into two halves, (i) efficient optimizers, which are among the top-performers, returning the lowest error among different molecular properties, and (ii) error-prone optimizers, which are among the lower-end performers. On the other hand, the performance of the optimizers in the noisy setting can be divided based on the resistance against the noise, (i) noise-resistant optimizers, returning a decent result even in the presence of noise, and (ii) noise-susceptible, which are the most affected by the noise.

\noindent While CG, SLSQP, and L\_BFGS\_S are efficient optimizers, they are highly noise susceptible, and while GD and ADAM are noise resistant, they are quite error-prone with SPSA being a close call of being error-prone. Ideally, one would employ those optimizers that can perform reasonably well in various situations, be it an ideal quantum scenario or a noisy quantum environment. From our analysis, two optimizers, POWELL and COBYLA, stand out as they perform reasonably well in ideal as well as noisy conditions. With some extra effort, as highlighted in~Section~\ref{sec:spsa}, SPSA could also well be an all-around optimizer. 

\subsection{Noise Resilience of SPSA, POWELL, and COBYLA Optimizers}
\noindent SPSA~\cite{spsa} is a gradient-based optimizer, which provides a technique to optimize systems with multiple unknown parameters, and it is particularly suited to large high-dimensional problems like large-scale population models, adaptive modeling, and simulation optimization. The highlight of the SPSA optimizer is the stochastic gradient approximation, which requires that the objective function be measured only twice, regardless of the size of the system or the dimension of the optimization problem. This results in a significant decrease in the optimization cost, especially in problems with a large number of variational parameters.\\
\noindent Let $H(\theta)$ be the loss function, where $\theta$ is a $n-$dimensional vector. The SPSA optimization generally has the form:
\begin{equation}
   \hat{\theta}_{k+1} = \hat{\theta}_{k} - a_k \hat{g}_{k}(\hat{\theta}_{k}) 
\end{equation}
\noindent where, $\hat{g}_{k}(\hat{\theta}_{k})$ is the estimate of the gradient $g(\theta) = \frac{\partial L}{\partial \theta}$.\\
If $y$ denotes the measurement of the loss function $H$ at any point, we can have two approximations, one-sided gradient approximation, involving $y(\hat{\theta}_{k})$, and $y(\hat{\theta}_{k} + \rm{perturbation})$, and two-sided approximations, involving $y(\hat{\theta}_{k})$, and $y(\hat{\theta}_{k} \pm {\rm perturbation})$. In the SPSA optimization, all elements of $\hat{\theta}_{k}$ are simultaneously and randomly perturbed to obtain two measurements of $y$. For the two-sided simultaneous perturbation, we have 
\begin{equation}
   \hat{g}_{ki}(\hat{\theta}_{k}) = \frac{\hat{y}(\hat{\theta}_{k} + c_k \Delta_k) - \hat{y}(\hat{\theta}_{k} - c_k \Delta_k)}{2c_k \Delta_{ki}} 
\end{equation}
\noindent where, $\Delta_k = (\Delta_{k1}, \Delta_{k2}, \cdots ,\Delta_{kp})^T$ is the user-specified $p-$dimensional random perturbation vector. This random perturbation used for the gradient approximation  might actually be the reason behind the overall success of the SPSA algorithm since the gradient is not evaluated at some fixed parameter values but rather at multiple random directions. This randomness in gradient evaluation potentially counters the overarching effect of quantum noise.\\

\noindent POWELL~\cite{powellopt} is a gradient-free optimizer, and it is one of the few methods which do not require the function being optimized to be differentiable, although the function must be real-valued. The algorithm requires an initial point and some directional vectors called the initial search vectors. A sequential one-dimensional minimization is then performed along each of these vectors. \\

\noindent 
Without loss of generality, consider a quadratic function that needs to be optimized,
\begin{equation}
   H(\Vec{\theta}) = \Vec{\theta}^T A \Vec{\theta} - 2\Vec{b}^T \Vec{\theta} + c 
\end{equation}
\noindent where, $A$ is positive definite symmetric, $ b \in R^n$. 
For a set of non-zero conjugate directions $\Vec{u}_1, \cdots, \Vec{u}_n$, the minimum of $f(\Vec{x})$ in the space spanned by $\left\{\Vec{u}_i\right\}$ can be found at the point $\displaystyle \sum_{i=1}^{n} \beta_i \Vec{u}_i $, where~\cite{brent_2002_algorithms}
\begin{equation}
   \beta_i = \frac{\Vec{u}_{i}^{T} \Vec{b}}{\Vec{u}_{i}^{T} A \Vec{u}_i}. 
\end{equation}
\noindent With the $\Vec{\theta}_0$ as the initial point and $\{\Vec{u}_i\}$ as the set of conjugate vectors, the POWELL algorithm proceeds as follows,
\begin{itemize}
    \item $\beta_i$ is computed for $i=1, \cdots ,n$, to minimize $H(\Vec{\theta}_{i-1} + \beta_i\Vec{u}_i)$, and  $\Vec{\theta}_i$ is defined  as  $\Vec{\theta}_{i-1} + \beta_i \Vec{u}_i$.
    \item $\Vec{u}_i$ is replaced by $\Vec{u}_{i+1}$ for $i=1, \cdots ,n-1$.
    \item $\Vec{u}_n$ is replaced by $\Vec{\theta}_n - \Vec{\theta}_0$.
    \item $\beta$ is computed to minimize $ H(\Vec{\theta}_0 + \beta \Vec{u}_n) $, and $\Vec{\theta}_0$ is replaced by by $\Vec{\theta}_0 + \beta \Vec{u}_n$.
    \item The process is repeated till some convergence criteria are met.
\end{itemize}

\noindent COBYLA is a gradient-free simplex method, where the constrained problem is approximated iteratively by linear programming problems\cite{cobyla}. Each iteration involves solving an approximate linear problem to obtain the next guess. 
To optimize the function $ H(\theta_i) $
within the COBYLA method, $ H(\theta_i) $ is evaluated at $ i = 0, 1, 2, \cdots , n $ vertices of the simplex, and the edges are interpolated by a unique linear polynomial $ L(\theta), \theta \in R $. For each iteration a trust region radius $\Delta > 0 $ that can be adjusted automatically is also required. 
The next (set of) variables is found by minimizing $ L(\theta) $ subject to
\begin{equation}
   || \theta - \theta^{'} || \leq \Delta 
\end{equation}
where, $ H(\theta^{'}) $ is the minimum value found so far.\\
In unconstrained COBYLA, we have
\begin{equation}
   \theta = \theta^{'} - \left(\frac{\Delta}{||\nabla L||}\right)\nabla L. 
\end{equation}
$ H(\theta) $ is then evaluated and the next simplex is formed by replacing one of the vertices of the old simplex with $\theta$.\\

\noindent Both POWELL and COBYLA do not require gradient evaluation. One might attribute that to the better performance of these methods in noisy situations. Like SPSA, these methods also evaluate the loss function at multiple parameter values, dictated by conjugate vectors in POWELL and the polynomial function in COBYLA. Therefore, the difference in architecture does make a significant difference, which is also apparent in the overall performance of these optimizers in the variational quantum eigensolvers.

\section{Conclusions}

\noindent Based on all these results and discussion, the following conclusions can be drawn out:
\begin{itemize}
    \item In the ideal case, for the accuracy of the ground-state energy evaluation, almost all the optimizers perform equally well, apart from SPSA, AQGD, and NFT. But based on the simulation time and iterations for convergence, L\_BFGS\_B, CG, and SLSQP are the best options among the gradient-based optimizers. At the same time, COBYLA and POWELL remain the best-performing gradient-free optimizers, with POWELL having a particular advantage in taking the smallest number of iterations to converge to an acceptable tolerance (FIG.~\ref{fig:iter}). Noise in quantum circuits presents a huge roadblock in practical applications of variational quantum algorithms in quantum chemistry as once the noise is introduced to the quantum circuits, the performance of many optimizers is significantly affected. However, when it comes to ground state energy error, almost all the optimizers perform similarly, with the error being of the same order across the board.
    \item For the correctness of the dissociation energy, one can see a more varied performance across different optimizers in both ideal and noisy simulators. Under ideal conditions, CG, SLSQP, and POWELL turned out to be the best-performing optimizers. It can be noted that L\_BFGS\_B performs exceptionally well in most cases, however, in Hydrogen Fluoride, it yields a large error in the dissociation energy. It may indicate the inefficiency of L\_BFGS\_B once the state space becomes larger. Under the noisy conditions, the performance of almost all the optimizers was affected, with SPSA, GD, POWELL, and COBYLA outperforming all others. 
    \item For the dipole moment, the trend remains the same with L\_BFGS\_B, CG, and SLSQP being the best gradient-based performers and COBYLA and POWELL being the most efficient gradient-free methods with ideal simulators. In the presence of noise, SPSA, POWELL, CG, and COBYLA outperform the other optimizers.  
\end{itemize}

\noindent In the ideal scenario, plenty of optimizers perform exceptionally well in variational quantum algorithms for chemistry applications with L\_BFGS\_B, CG, and SLSQP being the overall best-performing optimizers. However, in the present era of noisy devices, the options become somewhat limited with SPSA, POWELL, and COBYLA being the decent performers for the task at hand. The two most straightforward solutions for improving the performance of the classical optimizers in variational quantum eigensolvers are employing error correction methods to counter the noise in the system, or simply, waiting for better quality quantum systems to be available for use. Considering that the modern NISQ era is going to last for a while, one might also look out for a better class of optimizers. As we have encountered in this work, perturbative and gradient-free methods seem to have a clear edge over other techniques in noisy environments, so further investigation into these methods and tuning their hyperparameters for specific tasks can lead to some fruitful results.

\section*{Acknowledgements}

\noindent This work used the Supercomputing facility of IIT Kharagpur established under the National Supercomputing Mission (NSM), Government of India, and supported by the Centre for Development of Advanced Computing (CDAC), Pune. HS acknowledges the Ministry of Education, Govt. of India, for the Prime Minister's Research Fellowship (PMRF), Dr.~Varun Sharma (HMC Healthcare), and Mr.~Indranil Ghosh (Massey University) for helpful discussion. \\

\bibliography{main.bib}

\end{document}